\documentclass[useAMS,usenatbib]{mn2e}
\usepackage{multirow}
\usepackage{epsfig}
\usepackage{subfigure}

\makeatletter
\let\jnl@style=\rm
\def\ref@jnl#1{{\jnl@style#1}}

\def\aj{\ref@jnl{AJ}}                   
\def\araa{\ref@jnl{ARA\&A}}             
\def\apj{\ref@jnl{ApJ}}                 
\def\apjl{\ref@jnl{ApJ}}                
\def\apjs{\ref@jnl{ApJS}}               
\def\ao{\ref@jnl{Appl.~Opt.}}           
\def\apss{\ref@jnl{Ap\&SS}}             
\def\aap{\ref@jnl{A\&A}}                
\def\aapr{\ref@jnl{A\&A~Rev.}}          
\def\aaps{\ref@jnl{A\&AS}}              
\def\azh{\ref@jnl{AZh}}                 
\def\baas{\ref@jnl{BAAS}}               
\def\jrasc{\ref@jnl{JRASC}}             
\def\memras{\ref@jnl{MmRAS}}            
\def\mnras{\ref@jnl{MNRAS}}             
\def\pra{\ref@jnl{Phys.~Rev.~A}}        
\def\prb{\ref@jnl{Phys.~Rev.~B}}        
\def\prc{\ref@jnl{Phys.~Rev.~C}}        
\def\prd{\ref@jnl{Phys.~Rev.~D}}        
\def\pre{\ref@jnl{Phys.~Rev.~E}}        
\def\prl{\ref@jnl{Phys.~Rev.~Lett.}}    
\def\pasp{\ref@jnl{PASP}}               
\def\pasj{\ref@jnl{PASJ}}               
\def\qjras{\ref@jnl{QJRAS}}             
\def\skytel{\ref@jnl{S\&T}}             
\def\solphys{\ref@jnl{Sol.~Phys.}}      
\def\sovast{\ref@jnl{Soviet~Ast.}}      
\def\ssr{\ref@jnl{Space~Sci.~Rev.}}     
\def\zap{\ref@jnl{ZAp}}                 
\def\nat{\ref@jnl{Nature}}              
\def\iaucirc{\ref@jnl{IAU~Circ.}}       
\def\aplett{\ref@jnl{Astrophys.~Lett.}} 
\def\apspr{\ref@jnl{Astrophys.~Space~Phys.~Res.}}
\def\bain{\ref@jnl{Bull.~Astron.~Inst.~Netherlands}}
\def\fcp{\ref@jnl{Fund.~Cosmic~Phys.}}  
\def\gca{\ref@jnl{Geochim.~Cosmochim.~Acta}}   
\def\grl{\ref@jnl{Geophys.~Res.~Lett.}} 
\def\jcp{\ref@jnl{J.~Chem.~Phys.}}      
\def\jgr{\ref@jnl{J.~Geophys.~Res.}}    
\def\jqsrt{\ref@jnl{J.~Quant.~Spec.~Radiat.~Transf.}}
\def\memsai{\ref@jnl{Mem.~Soc.~Astron.~Italiana}}
\def\nphysa{\ref@jnl{Nucl.~Phys.~A}}   
\def\physrep{\ref@jnl{Phys.~Rep.}}   
\def\physscr{\ref@jnl{Phys.~Scr}}   
\def\planss{\ref@jnl{Planet.~Space~Sci.}}   
\def\procspie{\ref@jnl{Proc.~SPIE}}   

\makeatother

\title[A BLR origin for the iron K$\alpha$ line in NGC~7213]{A BLR origin for the iron K$\alpha$ line in NGC~7213}

\author[Stefano Bianchi, et al.]{Stefano Bianchi$^1$\thanks{E-mail: bianchi@fis.uniroma3.it (SB)}, Fabio La Franca$^1$, Giorgio Matt$^1$, Matteo Guainazzi$^2$, 
\newauthor
Elena Jimenez Bail\'on$^{3,4}$, Anna Lia Longinotti$^2$, Fabrizio Nicastro$^5$, Laura Pentericci$^5$\\
$^1$Dipartimento di Fisica, Universit\`a degli Studi Roma Tre, via della Vasca Navale 84, 00146 Roma, Italy\\
$^2$XMM-Newton Science Operations Center, European Space Astronomy Center, ESA, Apartado 50727, E-28080 Madrid, Spain\\
$^3$Instituto de Astronom\'ia, Universidad Nacional Aut\'onoma de M\'exico, Apartado Postal 70-264, 04510 Mexico DF, Mexico\\
$^4$LAEFF Apd. 78 Villanueva de la Ca\~nada - 28691-Madrid, Spain\\
$^5$Osservatorio Astronomico di Roma (INAF), Via Frascati 33, I-00040 Monte Porzio Catone, Italy\\
}

\begin{document}


\maketitle

\label{firstpage}

\begin{abstract}
The X-ray spectrum of NGC~7213 is known to present no evidence for Compton reflection, a unique result among bright Seyfert 1s. The observed neutral iron K$\alpha$ line, therefore, cannot be associated with a Compton-thick material, like the disc or the torus, but is due to Compton-thin gas, with the Broad Line Region (BLR) as the most likely candidate. To check this hypothesis, a long \textit{Chandra} HETG observation, together with a quasi-simultaneous optical spectroscopic observation at the \textit{ESO} NTT EMMI were performed. We found that the iron line is resolved with a FWHM=$2\,400^{+1\,100}_{-600}$ km s$^{-1}$, in perfect agreement with the value measured for the broad component of the H$\alpha$, $2640^{+110}_{-90}$ km s$^{-1}$. Therefore, NGC~7213 is the only Seyfert 1 galaxy whose iron K$\alpha$ line is unambiguously produced in the BLR. We also confirmed the presence of two ionised iron lines and studied them in greater detail than before. The resonant line is the dominant component in the Fe \textsc{xxv} triplet, therefore suggesting an origin in collisionally ionised gas. If this is the case, the blueshift of around $1\,000$ km s$^{-1}$ of the two ionised iron lines could be the first measure of the velocity of a starburst wind from its X-ray emission.
\end{abstract}

\begin{keywords}
galaxies: active - galaxies: Seyfert - X-rays: individual: NGC7213
\end{keywords}

\section{Introduction}

A narrow component of the iron K$\alpha$ line is almost invariably present in \textit{Chandra} high energy gratings and XMM-Newton CCD spectra of Active Galactic Nuclei \citep[e.g.][and references therein]{bianchi07,yaq04}. The line, typically unresolved with upper limits of several thousands km s$^{-1}$ for the Full Width at Half Maximum (FWHM), must be produced far from the nucleus, either in the torus envisaged in the Unification Model \citep{antonucci93} or in the Broad Line Region (BLR). In the former case, and if the torus is Compton-thick, a Compton reflection component should also be present. In the latter case, a much fainter Compton reflection component is expected, and the intrinsic width of the iron line should be the same as that of the optical broad lines. The torus hypothesis seems to be preferred for the majority of sources \citep[e.g.][]{bianchi04,nan06}, but to unambiguously distinguish between these hypotheses high energy resolution observations are required, which at present only
\textit{Chandra} can provide.

The Seyfert 1/LINER NGC~7213 (z=0.005839) presents a negligible amount of Compton reflection \citep[R=$\Delta\Omega/2\pi<0.19$:][]{bianchi03b,bianchi04}. This result, confirmed by three other BeppoSAX observations, is unique among bright Seyfert 1s observed by BeppoSAX \citep{per02,risa02,dad08}. Therefore, the observed neutral iron line, whose equivalent width is $\simeq80$ eV, cannot be associated with a Compton-thick material, like the disc or the torus, but is due to Compton-thin gas, like the BLR. The line width is unresolved by XMM-Newton EPIC pn, giving only an upper limit of about $8\,000$ km s$^{-1}$. In this Letter, we present a long \textit{Chandra} grating observation of the iron line width, which turns out to be resolved and fully consistent with the FWHM of the broad component of the H$\alpha$ line, as measured in a quasi-simultaneous optical observation.

\section{Data reduction}

\subsection{\label{xraydata}X-ray observation}

NGC~7213 was observed in two consecutive \textit{Chandra} Advanced CCD Imaging Spectrometer \citep[ACIS:][]{acis} observations, with the High Energy Transmission Grating Spectrometer \citep[HETGS:][]{hetg} in the focal plane. The first observation (obsid 7742) was performed on 2007, August 6th, for 115 ks, while the second one (obsid 8590) on 2007, August 9th, for 35 ks. Data were reduced with the Chandra Interactive Analysis of Observations \citep[CIAO:][]{ciao} 4.0.1 and the Chandra Calibration Data Base (CALDB) 3.4.2 software,
adopting standard procedures. First order High Energy Grating (HEG) and Medium Energy Grating (MEG) spectra were extracted for the source and the background. After having verified that there is no significant variability neither within any single observation nor between the two, we co-added them with the tool \textsc{add\_grating\_spectra}, in order to have a single HEG spectrum and a single MEG spectrum, for a total exposure time of 148 ks each.

All the fits were performed with \textsc{Xspec} 12.4.0 \citep{xspec}, taking into account the instrumental resolution with the response matrix computed with standard procedures. Local fits were performed around the iron K$\alpha$ band with the unbinned spectra, using the \citet{cash76} statistic, in order to take advantage of the high spectral resolution of the gratings. Broad band (0.4-10 keV) fits were instead performed on the binned spectra, where the large number of counts in each bin (250) allowed the use of the $\chi^2$ statistic. We did not analyse the 0th order data in this paper, since the count-rate per frame ($\simeq0.5$ s$^{-1}$ in the inner 2-pixel region) is larger than the one suggested for pileup-free spectra (see \textit{The Chandra ABC Guide to Pileup}\footnote{http://chandra.ledas.ac.uk/ciao4.0/download/doc/pileup\_abc.ps}), as confirmed by the resulting very flat spectrum.

\subsection{Optical observation}

In order to measure the physical width of the H$\alpha$ line, we asked for director discretionary time (ID: 279.B-5054A) to obtain a
quasi-simultaneous observation of NGC~7213 at the \textit{ESO} NTT telescope (La Silla, Chile). The observation was carried out in service mode on 2007, September 22th with the EMMI spectrograph equipped with grating \#7 and a 1-arcsec width slit. The wavelength resolution was 2.49 \AA\ (the pixel scale was 0.82 \AA) in the range 5650-7140 \AA. Four spectra, with 30 sec long integrations each, were obtained with 15 arcsec offsetted positions (pattern A B A B). The reduction process used standard \textsc{midas} and \textsc{iraf} facilities. The raw data were bias-subtracted, corrected for pixel-to-pixel variations (flat field) and eventually sky-subtracted. Wavelength calibrations were carried out by comparison with exposures of He and Ar lamps, with an accuracy of 0.17 \AA\ (1 $\sigma$). Relative flux calibration was carried out by observations of the spectrophotometric standard star LTT9239 \citep{hamuy92,hamuy94}. The final S/N ratio was about 22.

\section{Spectral analysis}

In the following, errors correspond to the 90\% confidence level for one interesting parameter ($\Delta \chi^2 =2.71$), while error bars in the plots correspond to 1 $\sigma$. Energies and wavelengths are reported in the rest-frame of the source. The Galactic column density along the line of sight to NGC~7213 is included \citep[$2.04\times10^{20}$ cm$^{-2}$: ][]{dl90}.

\subsection{The Fe complex}

\begin{figure}
\begin{center}
\epsfig{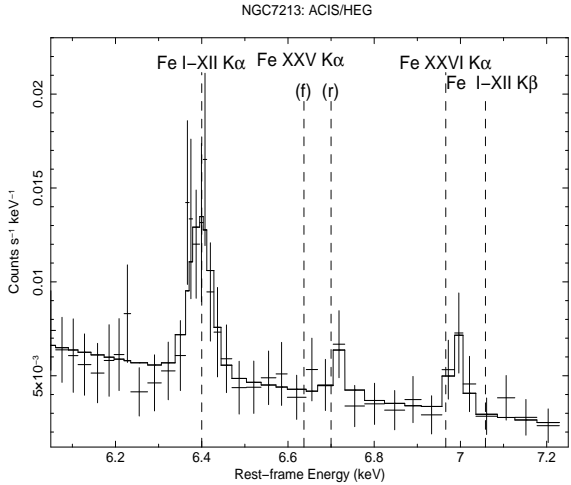}
\end{center}
\caption{\label{fekacomplex}\textit{Chandra} HEG spectrum of NGC~7213 around the iron line, along with the best fit with three Gaussian emission lines. The spectrum is re-binned for clarity and the positions of the most important transitions are marked.}
\end{figure}

Figure \ref{fekacomplex} shows the HEG spectrum in the 6.-7.5 keV band. Three significant emission lines are apparent. The strongest is the neutral K$\alpha$ line at $6.397^{+0.006}_{-0.011}$ keV, consistent with iron less ionised than Fe \textsc{xii} \citep{house69}. The observed flux ($2.9^{+0.9}_{-0.7}$ ph cm$^{-2}$ s$^{-1}$) and the EW ($120^{+40}_{-30}$ eV) are in agreement, within errors, with those found by XMM-\textit{Newton} \citep{bianchi03b}. The iron K$\alpha$ line is actually composed of a doublet, K$\alpha_1$ at 6.404 keV and K$\alpha_2$ at 6.391 keV, with a flux ratio 2:1 \citep{bea67}. Although the separation of this doublet cannot be resolved by the \textit{Chandra} HEG, we adopted a rigorous approach fitting two Gaussians, separated by 13 eV, the flux ratio fixed to the expected one and the width free to vary, but forced to be the same for the two lines. The width of the lines is resolved: $\sigma=22^{+10}_{-6}$ eV. Forcing the lines to be narrow ($\sigma=0$) leaves large residuals and a significantly worse fit ($\Delta\mathrm{C}=+15$). We note here that the measure of the width is completely unaffected by this modelling with respect to a single Gaussian, as already observed by \citet{yaq01} for other sources.

\begin{figure}
\begin{center}
\epsfig{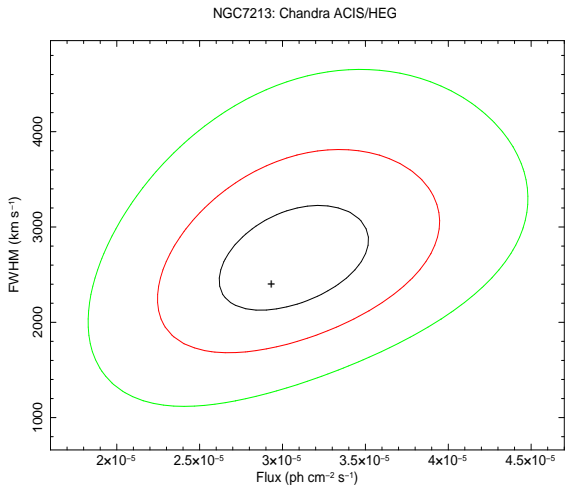}
\end{center}
\caption{\label{fekacontours}Flux vs. FWHM contour plot for the neutral iron line in NGC~7213. The curves refer to $\Delta$C = 2.30, 4.61 and 9.21, i.e. confidence levels of 68, 90 and 99\% for two interesting parameters.}
\end{figure}

Let us investigate the origin of the observed broadening. A possible origin for the observed width may be the presence of a Compton Shoulder (CS). If the CS is modelled with a Gaussian line with centroid energy at 6.3 keV and $\sigma=40$ eV \citep{matt02}, while the iron line width is fixed to 0, the fit is significantly worse ($\Delta\mathrm{C}=+14$) and only an upper limit on the CS flux is found. This is not surprising, given the absence of a Compton reflection component associated to the CS. Another explanation for the width of the iron line can be blending with iron lines from higher ionisation states, but this would imply a larger value for the centroid energy of the resulting line and a clear asymmetric profile. Therefore, Doppler broadening is left as the most likely explanation for the width of the iron line, which would correspond to a FWHM=$2\,400^{+1\,100}_{-600}$ km s$^{-1}$ (see Fig. \ref{fekacontours}).

Along with the neutral iron line, two other significant emission lines are observed in the HEG spectrum of NGC~7213, at $6.721^{+0.010}_{-0.016}$ and $6.987^{+0.019}_{-0.010}$ keV (see Fig. \ref{fekacomplex} and Table \ref{tableX}). The latter can be interpreted as emission from Fe \textsc{xxvi}, but its energy is blueshifted by a velocity of $900^{+800}_{-400}$ km s$^{-1}$ with respect to the theoretical value of 6.966 keV. If the other emission line is identified with the forbidden component of Fe \textsc{xxv} K$\alpha$ (6.637 keV), the blueshift must be $3\,800^{+400}_{-800}$ km s$^{-1}$. If instead it is identified with the resonant component (6.700 keV), its corresponding blueshift would be $900^{+500}_{-700}$ km s$^{-1}$, i.e. consistent with the one found for the Fe \textsc{xxvi} line. The fluxes and EWs of the two lines (see Table \ref{tableX}) are consistent with those found by XMM-\textit{Newton} \citep{bianchi03b}. Finally, no neutral K$\beta$ is found. The 90\% confidence level upper limit on its flux ($4\times10^{-6}$ ph cm$^{-2}$ s$^{-1}$) leads to an upper limit on the ratio against K$\alpha$ of 0.18, which is fully consistent with the expected value for neutral iron \citep[see][]{mbm03}.

\begin{table}
\caption{\label{tableX}NGC~7213: the iron line complex in the \textit{Chandra} HEG spectrum.}
\begin{center}
\begin{tabular}{cccccc}
Line & $\mathrm{E_T}$ &$\Delta E$ & FWHM & Flux & EW \\
(1) & (2) & (3) & (4) & (5) & (6)\\
\hline
 &  &  &  & &\\
Fe \textsc{i}-\textsc{xii} & 6.400$^*$ & $-3^{+6}_{-11}$ & $2400^{+1100}_{-600}$ & $2.9^{+0.9}_{-0.7}$ & $120^{+40}_{-30}$\\[1ex]
Fe \textsc{xxv} (r) & 6.700 & $+21^{+10}_{-16}$ & $<5500$ & $0.7\pm0.5$ & $24\pm17$\\[1ex]
Fe \textsc{xxvi} & 6.966$^*$ & $+21^{+19}_{-10}$ & $<2700$ & $1.3\pm0.6$ & $60\pm30$\\
 &  &  &  &\\
\hline
\end{tabular}
\end{center}
\textsc{Notes}-- Col. (1) Identification. Col. (2) Theoretical energy in keV \citep[NIST:][]{nist}. $^*$Weighted mean of the doublet. Col. (3) E$_\mathrm{obs}$-E$_\mathrm{T}$ in eV, after correction for z=0.005839. Col. (4) km s$^{-1}$ Col. (5) $10^{-5}$ ph cm$^{-2}$ s$^{-1}$. Col. (6) eV.
\end{table}

While a broadband X-ray analysis of the \textit{Chandra} spectrum is beyond the scopes of this Letter, we note here that the 2-10 keV binned spectrum can be modelled by a simple powerlaw with $\Gamma=1.69\pm0.01$ and a flux of $2.32\pm0.04\times10^{-11}$ erg cm$^{-2}$ s$^{-1}$, corresponding to a luminosity of $\simeq1.7\times10^{42}$ erg s$^{-1}$. These values are perfectly consistent with what found by XMM-\textit{Newton} 6 years before \citep{bianchi03b}. No other significant emission or absorption lines are found.

\subsection{The optical spectrum}

\begin{figure*}
\begin{center}
\epsfig{file=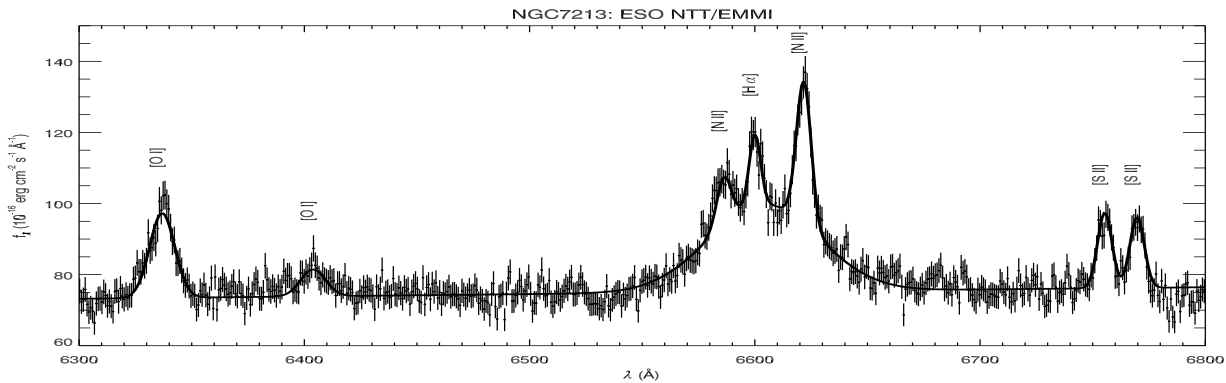, height=5.cm, width=17cm}
\end{center}
\caption{\label{opticalsp}\textit{ESO} NTT EMMI spectrum in the 6300 and 6800 \AA. The best fit is superimposed: the included lines are listed in Table \ref{tableopt}.}
\end{figure*}

The optical spectrum of NGC~7213 was analysed between 6300 and 6800 \AA. The most important atomic transitions expected in this wavelength range are clearly observed in the spectrum (see Fig. \ref{opticalsp}), so no attempt was made to remove the (unknown) contribution from the stellar continuum. Note that the optical spectrum analysed by \citet{fh84}, to whom we will refer, was not galaxy-subtracted longward of 5500 \AA, either.

\begin{figure}
\begin{center}
\epsfig{file=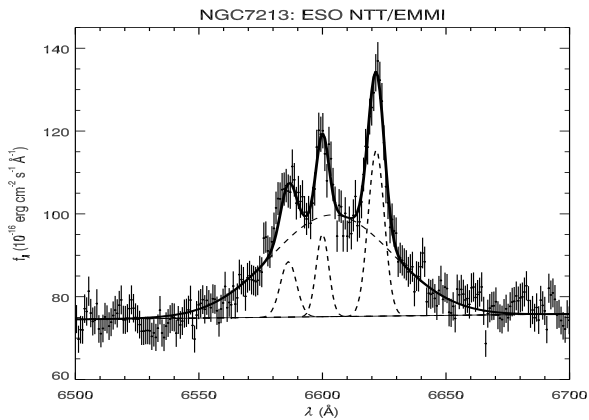, height=5.5cm, width=8cm}
\end{center}
\caption{\label{halpha}\textit{ESO} NTT EMMI spectrum of the H$\alpha$-{[{N\,\textsc{ii}}]} complex. The best fit is the result of the convolution of the four Gaussian lines plotted with broken lines and listed in Table \ref{tableopt}.}
\end{figure}

Fig. \ref{halpha} shows the H$\alpha$ blended with the {[{N\,\textsc{ii}}]} doublet. In order to disentangle any broad component of the H$\alpha$, we assumed that (i) \textit{F}($\mathrm{[{N\,\textsc{ii}}]}\, \lambda6583$/\textit{F}($\mathrm{[{N\,\textsc{ii}}]}\, \lambda6548$)=3, as required by the ratio of the respective Einstein coefficients, (ii) $\lambda_2/\lambda_1=6583.39/6548.06$ and (iii) the {[{N\,\textsc{ii}}]} lines are Gaussians with the same width. With these physical assumptions, a fit with three narrow emission lines (H$\alpha$ plus the {[{N\,\textsc{ii}}]} doublet) is very poor and strongly suggests the presence of a broad component. The addition of this further component leads to a very good fit: see Table \ref{tableopt} and Fig. \ref{halpha} for the 4 Gaussian lines included in the fit and their convolution which nicely matches the observed profile.

\begin{table}
\caption{\label{tableopt}NGC~7213: optical emission lines in the \textit{ESO} NTT EMMI spectrum.}
\begin{center}
\begin{tabular}{lllll}
Line & $\lambda_\mathrm{l}$ & $\Delta\lambda$ & FWHM & Flux \\
(1) & (2) & (3) & (4) & (5)\\
\hline
 &  &  &  &\\
$\mathrm{[{O\,\textsc{i}}]}$ & 6300.32 & -0.3 & $620^{+60}_{-70}$ & $3.38\pm0.01$\\[1ex]
$\mathrm{[{O\,\textsc{i}}]}$ & 6363.81 & +2.8 & $620^{+60}_{-70}$ & $0.94\pm0.01$\\[1ex]
H$\alpha$ & 6562.79 & -1.1 & $280^{+50}_{-30}$ & $1.33^{+0.13}_{-0.15}$ \\[1ex]
H$\alpha$ & 6562.79 & +2.6 & $2640^{+110}_{-90}$ & $14.75^{+0.18}_{-0.60}$ \\ [1ex]
$\mathrm{[{N\,\textsc{ii}}]}$ & 6548.06 & -0.1 & $370\pm30$ & $1.15\pm0.07$\\[1ex]
$\mathrm{[{N\,\textsc{ii}}]}$ & 6583.39 & -0.1 & $370\pm30$ & $3.45\pm0.21$\\[1ex]
$\mathrm{[{S\,\textsc{ii}}]}$ & 6716.42 & -0.2 & $350^{+30}_{-15}$ & $1.783^{+0.012}_{-0.016}$\\[1ex]
$\mathrm{[{S\,\textsc{ii}}]}$ & 6730.78 & -0.3 & $350^{+30}_{-15}$ & $1.773^{+0.011}_{-0.016}$\\
 &  &  &  &\\
\hline
\end{tabular}
\end{center}
\textsc{Notes.}-- Col. (1) Identification. Col (2) Laboratory wavelength (\AA), in air \citep{bowen60}. Col. (3) $\lambda_\mathrm{obs}-\lambda_\mathrm{l}$, after correction for z=0.005839. Typical error $\simeq\pm0.5$ \AA. Col. (4) km s$^{-1}$ (instrumental resolution not removed). Col. (5) $10^{-14}$ erg cm$^{-2}$ s$^{-1}$.

\end{table}

The widths of the narrow H$\alpha$ and the {[{N\,\textsc{ii}}]} doublet are similar (see Table \ref{tableopt}), while the FWHM of the broad component of H$\alpha$ ($2640^{+110}_{-90}$ km s$^{-1}$) clearly requires a different origin, i.e. the BLR. Our results are qualitatively in agreement with those of \citet{fh84}, in the sense that at least a broad component for the H$\alpha$ is required by the data (they actually fit a total of seven components, but with the warning that the physical significance of all these lines is questionable). Moreover, the FWHM we derive for the H$\alpha$ is close to the width of the broad component of the H$\beta$ as measured by \citet{fh84} and \citet{win92}, around $4\,000$ and $3\,200$ km s$^{-1}$, respectively.

Finally, the {[{S\,\textsc{ii}}]} doublet has a FWHM consistent with that of the {[{N\,\textsc{ii}}]} doublet and the narrow H$\alpha$ component, while the {[{O\,\textsc{i}}]} doublet is broader, but all in agreement with an origin in the Narrow Line Region.

\section{Discussion}

Differently from all the other Seyfert 1s with broadband X-ray observations, NGC~7213 lacks a Compton reflection component, the clearest signature for the presence of a Compton-thick material. Therefore, the observed neutral iron line must originate in a Compton-thin material, be it the BLR or a Compton-thin torus.

The \textit{Chandra} HEG spectrum of NGC~7213 allowed us to resolve, for the first time, the iron line width at $2\,400^{+1\,100}_{-600}$ km s$^{-1}$ (FWHM). The analysis of a quasi-simultaneous optical observation confirmed the presence of a broad component of the H$\alpha$ line, for which we measured a FWHM=$2640^{+110}_{-90}$ km s$^{-1}$. The widths of the two lines are in very good agreement, which suggests that they are likely to be produced in the same material.

To test if this scenario is possible, we have to verify if the observed EW of the iron line is in agreement with an origin in the BLR. Following the detailed procedure described by \citet{yaq01}, derived from \citet{kk87} adopting updated atomic data, the expected EW for the BLR in NGC~7213 can be written as:

\begin{equation}
\label{formulaew}
\mathrm{EW_{FeI}}\simeq34\,\left(\frac{f_c}{0.35}\right)\left(\frac{N_\mathrm{H}}{10^{23}\,\mathrm{cm}^{-2}}\right)\,\mathrm{eV}
\end{equation} 

This estimate assumes a spherically symmetric cloud distribution for the BLR, with a covering factor of $f_c$, a column density of $N_\mathrm{H}$ for each cloud and an iron abundance relative to hydrogen of $A_\mathrm{Fe}=4.68\times10^{-5}$ \citep{ag89}. The powerlaw index of $\Gamma=1.69$ derived from the \textit{Chandra} and the XMM-\textit{Newton} observations was further adopted in the derivation of (\ref{formulaew}). Assuming $f_c=0.35$, a column density around $3\times10^{23}$ cm$^{-2}$ can reproduce an EW$\simeq100$ eV, which is the order of magnitude found by \textit{Chandra} and XMM-\textit{Newton}. These values for $f_c$ and $N_\mathrm{H}$ are within the ranges usually assumed in photoionization models of BLRs \citep{netzer90}. In particular, $f_c=0.35$ is the value derived by \citet{gk98} from observational constraints in the Seyfert 1 NGC~5548, even if more `canonical' values around 0.1 and 0.25 are generally found. The large covering factor needed to account for the iron line EW may be at odds with the regular intensity of the observed broad optical lines, but the weakness or absence of an UV bump in NGC~7213 \citep[e.g.][]{starl05} may result in a deficit of UV photoionizing photons and compensate the geometrical factor. In any case, $f_c$ and $N_\mathrm{H}$ can be lower, provided that $A_\mathrm{Fe}$ is larger than solar and/or the X-ray illumination of the BLR is anisotropic \citep[see e.g.][and references therein]{yaq01}. On the other hand, we cannot exclude a further contribution to the iron line EW from a Compton-thin torus, located on a pc-scale, but we stress here that we do not have any other piece of evidence to support its presence.

The presence of emission lines from highly ionised iron in the \textit{Chandra} HEG spectrum of NGC~7213 confirms the XMM-\textit{Newton} results by \citet{bianchi03b}. As suggested by these authors, their origin may be in gas photoionised by the AGN, as found in many Seyfert 1s and 2s \citep[e.g.][]{bm02,bianchi05}. However, the gratings data presented here revealed that the resonant line is the dominant component in the Fe \textsc{xxv} triplet (unless its blueshift is significantly larger than the one measured for the Fe \textsc{xxvi} line), which is suggestive of an origin in gas in collisional equilibrium. This is in agreement with the results found by \citet{starl05} using diagnostic tools based on the O \textsc{vii} triplet observed in the XMM-\textit{Newton} RGS spectrum of NGC~7213. It is interesting to note that the Fe \textsc{xxv}-\textsc{xxvi} lines' blueshift of around $1\,000$ km s$^{-1}$ \citep[consistent with the centroid energies of the O \textsc{vii} lines as measured by][]{starl05}, together with the evidence that the gas producing such lines is in collisional ionization equilibrium, suggest that the hot, line emitting, gas be in the form of a starburst driven wind \citep[see e.g.][and references therein]{heck03}, rather than an intrinsic (typically photoionized) AGN wind.

The H$\alpha$ morphology of NGC~7213 was studied in detail by \citet{sb96} and \citet{hameed01}. A circumnuclear ring of star formation is located at 20-30 arcsec from the nucleus (corresponding to a few kpc) and a giant H$\alpha$ filament is present at 2.9 arcmin of distance (around 19 kpc). Neither of these structures is observed in the \textit{Chandra} X-ray image, which only shows an unresolved (within few arcsec) nucleus. However, \citet{sb96} reported the possible presence of a collimated outflow within 14 arcsec from the nucleus, with a velocity around 100 km s$^{-1}$. The gas producing the ionised iron lines reported in this Letter may be the inner, hotter and faster phase of this superwind. If this is the case, this could be the first measure of the velocity of a starburst wind from its X-ray emission.

\section*{Acknowledgements}

Based on observations made using Director Discretionary Time at the NTT/ESO telescope at the La Silla observatory under program ID:
279.B-5054A. SB and GM acknowledge financial support from ASI (grant 1/023/05/0). F.N. acknowledges support from NASA grant GO7-8095X. We thank F.~J. Carrera for his help on analysing optical data and the referee for useful suggestions.

\bibliographystyle{mn}
\bibliography{sbs}

\label{lastpage}

\end{document}